\documentclass[aps,prb,superscriptaddress,reprint,showpacs,floatfix]{revtex4-1}

\usepackage{tabularx}
\usepackage{float}
\usepackage{soul}
\usepackage{comment}
\usepackage{bm}
\usepackage{graphicx}
\usepackage{amsmath}
\usepackage{mathtools}
\usepackage{xcolor}
\usepackage{mathrsfs}
\usepackage[colorlinks=true,citecolor=blue]{hyperref}
\hypersetup{colorlinks=true,citecolor=blue,linkcolor=blue,urlcolor=blue}

\usepackage{chemformula} % Formula subscripts using \ch{}
\usepackage[T1]{fontenc} % Use modern font encodings
\usepackage{physics}

\usepackage{chngcntr}

\begin{document}

\title{Evidence for Phonon-Assisted Intertube Electronic Transport in an Armchair Carbon Nanotube Film}

\author{Davoud~Adinehloo}
\affiliation{Department of Electrical Engineering, University at Buffalo, Buffalo, NY 14228, USA}

\author{Weilu~Gao}
\affiliation{Department of Electrical and Computer Engineering, University of Utah, Salt Lake City, UT 84112, USA}

\author{Ali~Mojibpour}
\affiliation{Department of Electrical and Computer Engineering, Rice University, Houston, TX 77005, USA}

\author{Junichiro~Kono}
\affiliation{Department of Electrical and Computer Engineering, Rice University, Houston, TX 77005, USA}
\affiliation{Department of Physics and Astronomy, Rice University, Houston, TX 77005, USA}
\affiliation{Department of Materials Science and NanoEngineering, Rice University, Houston, TX 77005, USA}
\affiliation{The Smalley-Curl Institute, Rice University, Houston, TX 77005, USA}

\author{Vasili~Perebeinos}
\affiliation{Department of Electrical Engineering, University at Buffalo, Buffalo, NY 14228, USA}
\email{vasilipe@buffalo.edu}

\begin{abstract}
The electrical conductivity of a macroscopic assembly of nanomaterials is determined through a complex interplay of electronic transport within and between constituent nano-objects.  Phonons play dual roles in this situation: their increased populations tend to reduce the conductivity via electron scattering while they can boost the conductivity by assisting electrons to propagate through the potential-energy landscape. Here, we identify a  phonon-assisted coherent electron transport process between neighboring nanotubes in temperature-dependent conductivity measurements on a macroscopic film of armchair single-wall carbon nanotubes.  Through atomistic modeling of electronic states and calculations of both electronic and phonon-assisted junction conductances, we conclude that phonon-assisted conductance is the dominant mechanism for the observed high-temperature transport. The unambiguous manifestation of coherent intertube dynamics proves a single-chirality armchair nanotube film to be a unique macroscopic solid-state ensemble of nano-objects promising for the development of room-temperature coherent electronic devices.
\end{abstract}

\maketitle
%\section{Introduction}

Toward large-scale applications of nanomaterials in electronics~\cite{Hersam2013,Yanagi2010,Hart2018,GaoetAl20JPD}, understanding and controlling electron transport processes, not only within each nano-object but also between them, is crucial~\cite{Zorn2021}. The overall electrical conductivity of a macroscopic assembly of nanomaterials is determined by an array of interdependent quantities – e.g., defect density, doping level, Fermi energy, material size, the density of nano-objects, purity, homogeneity, morphology, and temperature. The role of phonons in this highly complex situation is subtle since they tend to scatter electrons to reduce the conductivity while simultaneously assisting electrons to go through the potential-energy landscape to increase the conductivity. This dual role of phonons in electronic transport across macro-objects has not been elucidated. Particularly, phonon-assisted processes, including phonon-assisted coherent electron transfer between nano-objects, have not been identified in a macroscopic sample.

\begin{figure*}[htb]
\includegraphics[width=0.9\textwidth]{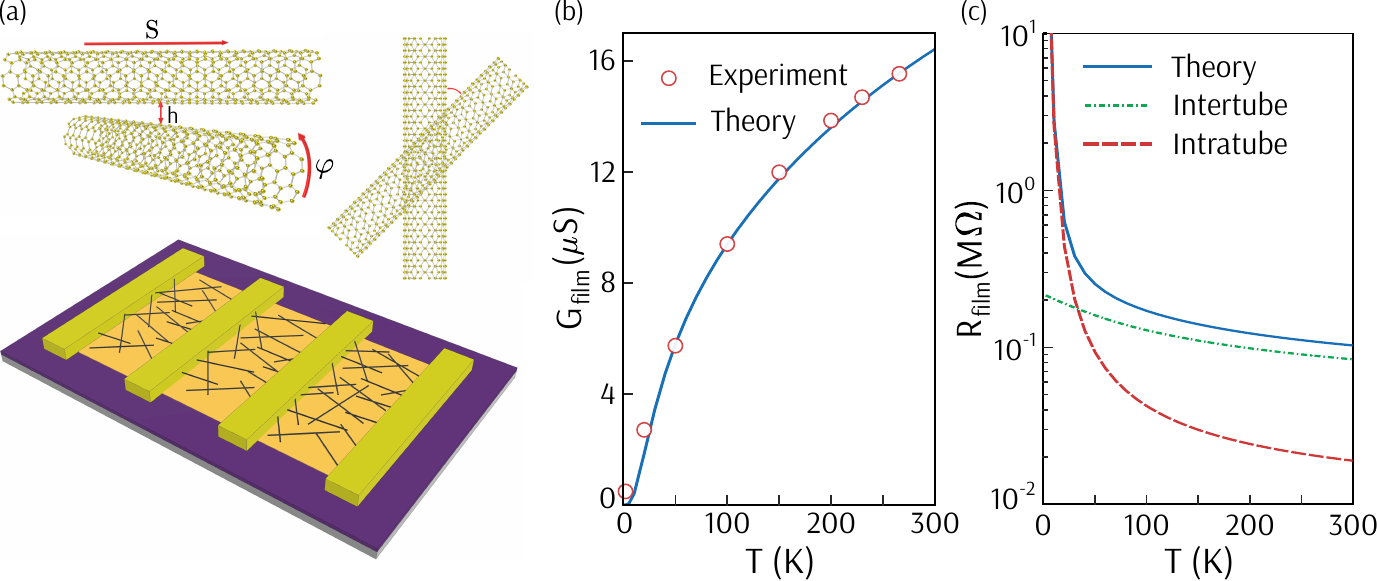}% Here is how to import EPS art
\caption{\label{fig:Fig1} (a)~Schematic illustration of the device used in this study. $\theta$ - crossing angle, $\varphi$ - rotating angle, $h$ - interlayer distance, and $S$ - sliding shift between CNTs.  (b)~Experimentally measured (red circles) and theoretically calculated (blue solid) film conductance of the (6,6) CNT film as a function of temperature. (c)~Theoretically obtained total resistance of the film (blue solid) and the decomposed contributions: intratube (red dashed) and a sum of the two electronic and phonon-assisted contributions to the intertube junction resistance (green dotted dashed). The calculations were performed at Fermi energy $E_\text{F}=10$\,meV and applied bias across the junction $V=1$\,mV.}
\end{figure*}

Here we study macroscopic assemblies of carbon nanotubes (CNTs), which provide an ideal model system in which to address the above issues and questions.  Since their discovery in the early 1990s,\cite{Iijima91Nature,IijimaIchihashi93Nature,BethuneetAl93Nature} extensive studies have revealed and established the truly unique electronic and optical properties of these one-dimensional nano-objects, particularly on individual nanotube levels.
Depending on the precise atomic arrangements of $sp^2$-bonded carbon atoms in the honeycomb lattice, specified by a pair of integers called chirality indices, ($n$,$m$), both metallic and semiconducting species of CNTs exist; details of the band structure, most importantly the band gap, are determined by ($n$,$m$).\cite{DresselhausetAl01Book,AvourisetAl07NN,AvourisetAl08NP,NanotetAl13Book,WeismanKono19Book}
Intertube electronic transport has also been studied theoretically\cite{nakanishi2001conductance,buldum2001contact,maarouf2011low} and experimentally.\citep{fuhrer2000crossed,FuhrerPhysicaE2000} Based on the momentum-matching conditions on the initial and final states of electrons in a transfer between two individual CNTs with a crossing angle $\theta$ (see Figure~\ref{fig:Fig1}a), the existence of special values of $\theta$, where the intertube conductance becomes  maximum  and minimum has been identified~\cite{nakanishi2001conductance,buldum2001contact,maarouf2011low}.
We show that after structural relaxation in a CNT- crossing geometry, the overall junction conductance increases by an order of magnitude with a quantitatively similar crossing angle dependence. Moreover, following Reference~\citenum{perebeinos2012phonon}  for twisted bilayer phonon-assisted junction conductance, we demonstrate here that the intertube phonon-assisted junction conductance at room temperature is comparable to the pure electronic conductance. At some crossing angles, phonon-assisted junction conductance exceeds its electronic counterpart by an order of magnitude. These two effects -- geometry relaxation and the additional phonon-assisted mechanism -- enable us to achieve better quantitative agreement with prior experiments\cite{fuhrer2000crossed} as well as the data reported here.

In our previous work,\citep{GaoetAl21Carbon} we investigated the temperature dependence of conductivity in a set of macroscopic single-chirality CNT films with well-defined ($n$,$m$).  Distinctly ($n$,$m$)-dependent and strongly temperature-dependent conductivity was observed, and the overall behaviors were explained through the Mott variable range hopping (VRH) model in a wide temperature range. However, one of the samples -- an armchair (6,6) CNT film -- exhibited clear deviation from the VRH behavior, especially at elevated temperatures, indicating that a different transport mechanism was at work. Through detailed quantitative analysis of the temperature-dependent conductivity, we found that the localization length was longer in (6,6) CNT film than in the other chirality samples such that intertube transport, as opposed to intratube transport, dominated in the (6,6) film at high temperatures. Here, through modeling and calculations of both electronic and phonon-assisted junction conductances, we unambiguously identify phonon-assisted conduction as the dominant mechanism for the observed high-temperature transport behaviour. The clear manifestation of this coherent dynamic process makes a single-chirality armchair CNT film a unique macroscopic object in which to study quantum transport processes at room temperature.

The device we used in this study was based on a thin film of randomly oriented, chirality-enriched (6,6) CNTs.  The device had four electrodes, and the channel between the two inner electrodes, shown in the bottom panel of Figure~\ref{fig:Fig1}a, had a length ($L$), width ($W$), and thickness ($t$) of 8\,$\mu$m, 5\,$\mu$m, and $\sim$12\,nm, respectively.  Further experimental details are fully described in Reference~\citenum{GaoetAl21Carbon}. The experimentally measured film conductance, $G_{\rm film}$, which is equal to the film conductivity times $Wt/L$, is plotted against temperature in Figure~\ref{fig:Fig1}b, showing that $G_{\rm film}$ monotonically but sublinearly increases with temperature as the temperature is raised from 4.2\,K to 300\,K.

To understand the observed temperature dependence of the film conductance, we consider a series resistance model of the CNT conductivity and junction resistance, which contribute to the overall resistance of the film as\cite{shim2015optimally}:
\begin{equation}\label{eq:Gfilm}
    G_\text{film}^{-1} = C \left (\frac{L_\text{cnt}}{\sigma_\text{intra}} + \frac{1}{G_\text{inter}} \right ),
\end{equation}
where the first and second terms originate from the intratube conductivity $\sigma_\text{intra}$ and the intertube junction conductance $G_\text{inter}$, respectively. A geometrical prefactor, $C= L(d+h)/(tW)$, is determined by the average number of junctions in the film, where $d$ is the CNT diameter, $h$ is the interlayer distance between CNTs, and $L_\text{cnt}$ is the average length of CNTs.
If we assume that the average crossing angle in a randomly oriented CNT film is $\theta=45^{\circ}$, $C$ should be $0.16$, according to our geometry, while the best fit to the data was obtained for $C=0.09$, as shown in Figure~\ref{fig:Fig1}b.

A temperature-dependent film conductance can arise from various sources.  Our previous study on single-chirality CNT films\cite{GaoetAl21Carbon} has shown that VRH is the dominant mechanism at low temperatures. Therefore, we model $\sigma_\text{intra}$ by the 1D Mott VRH model\citep{GaoetAl21Carbon}, i.e., $\sigma_\text{intra} =\sigma_0 \exp[-(T_0/T)^{1/2}]$.
In Figure~\ref{fig:Fig1}b, the theory curve according to Equation~(\ref{eq:Gfilm}) includes the electronic and phonon-assistant contributions to the intertube (junction) conductance (discussed below) as well as the intratube conductivity. Since the sample is randomly oriented and the junction conductance depends on the angle $\theta$ (see Figure~\ref{fig:Fig1}a), we averaged the intertube junction conductance over $\theta$ between $20^{\circ}$ and $70^{\circ}$. The relative contributions of the intratube and intertube resistances are shown in Figure~\ref{fig:Fig1}c. At temperatures below 50\,K, the intratube resistance dominates, whereas, at temperatures above 50\,K, the intertube junction resistance gives a dominant contribution to the film resistance.

\begin{figure*}[htb]
\includegraphics[width=1.0\textwidth]{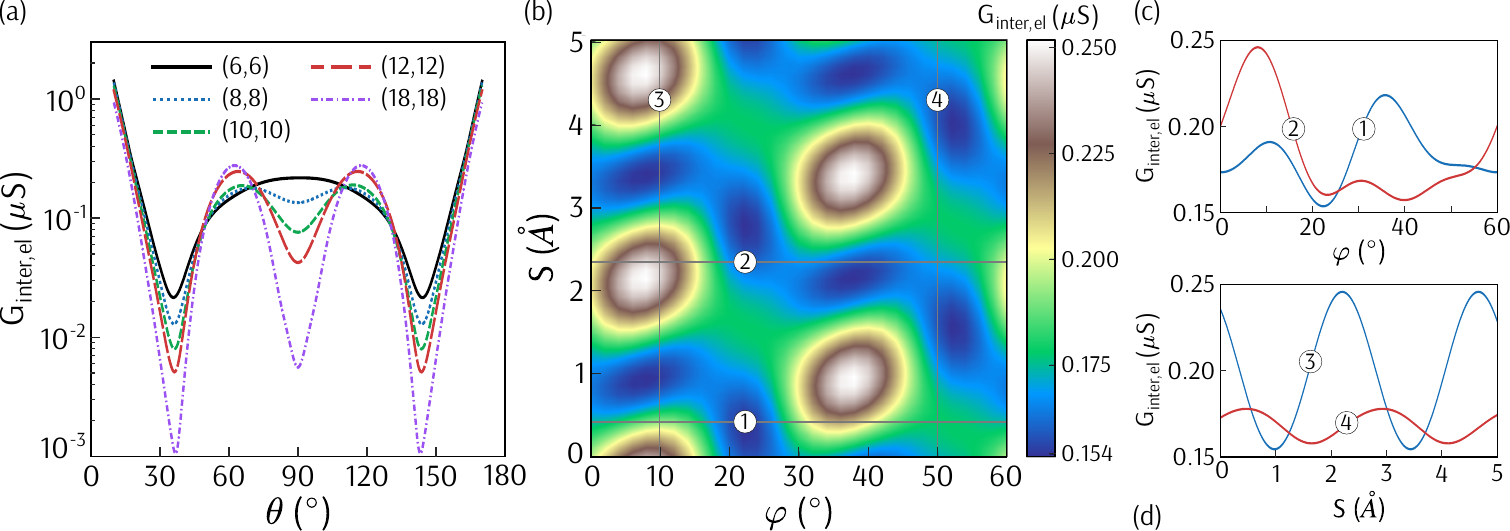}% Here is how to import EPS art
\caption{\label{fig:Fig2} (a)~Electronic conductance, $G_\text{inter, el}$, of crossed CNTs as a function of the crossing angle, $\theta$, between armchair CNTs of different diameters. (b)~Surface plot of $G_\text{inter, el}$ versus $S$ and $\varphi$ in CNTs. $G_\text{inter, el}$ versus $\varphi$ in (c) and versus $S$ in (d) at  horizontal and vertical cuts depicted in (b). In (b), (c), and (d), we used $\theta=70^{\circ}$ and (6,6) CNTs. All results were calculated at room temperature, $E_\text{F} = 0$ eV, and $V=1$\,mV.}
\end{figure*}

Below we discuss calculations of the intertube junction conductance. In the case of an ideally transparent contact, the maximum intertube conductance equals $4e^2/h\approx155$ $\mu$S. However, an experimentally measured conductance\cite{fuhrer2000crossed} is substantially smaller; therefore, we use perturbation theory (see Methods) to calculate the intertube junction conductance as a sum of two parallel channels: electronic conductance $G_{\rm inter, el}$ due to the overlap of $\pi$ orbitals on neighboring tubes described by the tight-binding model and phonon-assistant conductance $G_{\rm inter, ph}$ (see Methods).
Since the distance between $\pi$-orbitals in neighboring CNTs determines the hopping overlap, intertube junction conductance varies with the CNT structure registry.
Atomic positions in an unrelaxed structure depend on the crossing angle ($\theta$), rotation angle ($\varphi$), and sliding shift ($S$) between adjacent CNTs; see Figure~\ref{fig:Fig1}a. In Figure~\ref{fig:Fig2}a we show $G_\text{inter, el}$ for an unrelaxed structure versus crossing angle $\theta$ in different armchair CNTs, including (6,6) CNTs; $G_{\rm inter, el}$ can vary by more than two orders of magnitude as $\theta$ changes.
Figures~\ref{fig:Fig2}b--\ref{fig:Fig2}d show that $G_{\rm inter, el}$ varies with $\varphi$ and $S$ by about 30\% in (6,6) CNTs. Figure~\ref{fig:Fig2}b is a surface map of $G_{\rm inter, el}$  versus $\varphi$ and $S$ with $\theta=70^{\circ}$, while Figures~\ref{fig:Fig2}c and \ref{fig:Fig2}d are detailed $\varphi$ and $S$ dependences of $G_{\rm inter, el}$, corresponding to the labeled solid lines ($\o1-\o4$) in Figure~\ref{fig:Fig2}b. The variation of $G_{\rm inter, el}$ with $\varphi$ and $S$ are periodic with periods of $60^{\circ}$ and 2.46\,\AA, respectively, and hence, $G_{\rm inter, el}$ in Figure~\ref{fig:Fig2}a was obtained as an average over $\varphi$ and $S$.

\begin{figure*}
\includegraphics[width=0.9\textwidth]{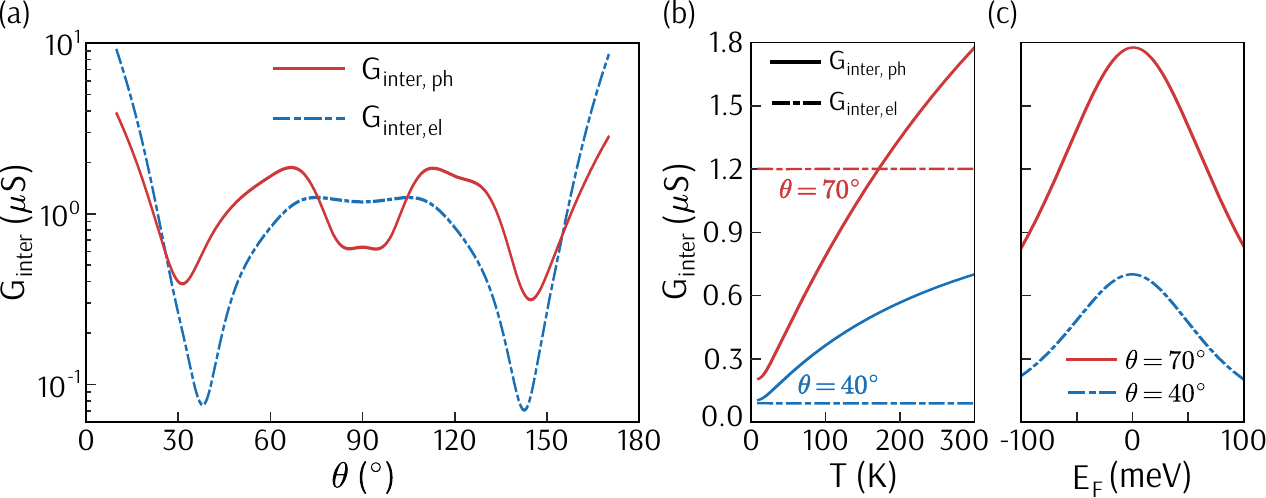}% Here is how to import EPS art
\caption{\label{fig:Fig3}(a)~Electronic (dashed blue) and phonon-assisted (solid red) intertube junction-conductance of relaxed (6,6) CNTs as a function of $\theta$ at room temperature. (b)~Electronic (dashed) and phonon-assisted (solid) intertube junction conductance of crossed CNTs for $\theta= 40^{\circ}$ (blue) and $\theta= 70^{\circ}$ (red).  (c)~shows phonon-assisted intertube junction conductance as a function of Fermi energy for $\theta= 40^{\circ}$ (dashed blue) and $\theta=70^{\circ}$ (solid red). We used a bias of $V=1$ mV in all panels and zero Fermi energy in panels (a) and (b).}
\end{figure*}

The large-diameter limit of $G_{\rm inter}$ can be understood by considering the junction conductance between twisted graphene bilayers,\cite{maarouf2011low} where the momentum conservation law of electrons in two Brillouin zones twisted with respect to each other by angle~$\theta$ governs the tunneling probability. However, in smaller-diameter CNTs, finite curvature plays an important role such that the finite contact area between the CNTs breaks the translation symmetry and helps relax the momentum conservation requirement. As a result, the conductance variations with angles diminish in magnitude, and a local minimum at $90^{\circ}$ vanishes in small-diameter armchair CNTs, as shown in Figure~\ref{fig:Fig2}a.

To better simulate the experimental situation, we relaxed the atomic positions at the junction using an atomistic valence force model\cite{perebeinos2009valence} for the internal distortions in each CNT and the Lennard-Jones potential for atoms on adjacent CNTs. We set the van der Waals adhesion energy to $E_\text{a}=60$ meV per atom\cite{zacharia2004interlayer} and the interlayer graphene bilayer spacing to $h = 3.35$\,\AA. The $G_{\rm inter, el}$ of the relaxed structure is increased by an order of magnitude compared to the unrelaxed case; this arises from the fact that more atoms on neighboring CNTs get closer due to CNT bending in the contact area.
Figure~\ref{fig:Fig3}a depicts the $G_{\rm inter, el}$ and $G_{\rm inter, ph}$ of the relaxed structure versus $\theta$, showing that overall they are comparable in magnitude at room temperature.
The $\theta$ dependence of $G_{\rm inter, ph}$ is weaker than that of $G_{\rm inter, el}$ because phonons help relax the momentum conservation law. Note also that the $S$ and $\varphi$ dependences of the conductance do not enter into consideration for the relaxed structure.

The temperature dependence of phonon-assisted conductance $G_\text{inter, ph}$ is nearly linear in Figure~\ref{fig:Fig3}b, reflecting the fact that low energy phonons are responsible for the phonon-assisted conductance, while the electronic intertube conductance is independent of temperature. The crossover temperature at which the two contributions become equal depends on the crossing angle. At angles near $\theta\sim30^{\circ}$, the phonon-assisted contribution dominates the intertube conductance for almost all temperatures, whereas at $\theta\sim90^{\circ}$ the electronic contribution  is larger even at room temperature, as shown in Figures~\ref{fig:Fig3}a and b.

Finally, Figure~\ref{fig:Fig3}c show the calculated Fermi energy dependence of $G_{\rm inter, ph}$.  A small variation of the  $G_{\rm inter, ph}$ with the Fermi energy in Figure~\ref{fig:Fig3}c reflects the fact that electron-phonon matrix elements have a weak momentum dependence and densities of electronic states are constant. In the case of electronic intertube conductance, we find that $G_{\rm inter, el}$ is almost independent of the Fermi energy (not shown).

In summary, we demonstrated that the temperature dependence of CNT film conductivity depend both on intratube and intertube (or junction) electronic transport. The dominant conduction mechanism depends on the film preparation conditions and the electronic structure of the CNTs.\cite{GaoetAl21Carbon} In this study, we showed that in films with a long localization length, such as a chirality-enriched (6,6) CNT film, the temperature dependence of the junction conductance can give rise to the observed overall temperature dependence of the CNT film conductance over a wide temperature range. We evaluated the temperature-independent electronic junction conductance and the temperature-dependent phonon-assisted junction conductance using microscopic theory, as shown in Figure~\ref{fig:Fig3}. Our numerical results show that both contributions are important at room temperature, and depending on the crossing angle, the phonon-assisted junction conductance can be significantly larger than the electronic intertube conductance. After averaging over the CNT crossing angle, our model accurately reproduces the experimental data over a wide temperature range with a single parameter fit known from the film device dimensions. We found that structural relaxation increases the electronic conductance by an order of magnitude, making calculated results more consistent with prior experimental results of junction conductance at low temperatures.\cite{fuhrer2000crossed} While the overall qualitative agreement between our model and the measured film conductance is evident, the nearly perfect quantitative agreement is likely accidental because of the oversimplified model used to estimate the numerical value of the parameter $C$ in Equation~(\ref{eq:Gfilm}). At the same time, we underestimate by a factor of three the intertube conductance measured on individual CNTs,\cite{fuhrer2000crossed} which could be due to a lack of angle control in the experiment. However, there can be a more fundamental reason for the observed discrepancy: the role of short-range defects in CNTs on the electronic and phonon-assistant intertube junction conductance, which should be a subject of further theoretical and experimental studies. Our work on chirality-enriched films provides guidelines for reinterpretation and analysis of other CNT network film resistances with the interplay of CNTs intratube transport and intertube junction transport.

\section{Methods}
\paragraph{\label{Theory:Gfilm}Conductance of 2D film:}As the thickness of the film is much smaller than the averaged length of SWCNT ($L_\text{cnt}$), the film can be considered to be a 2D film. Then, the number of junctions for each percolation path can be estimated by $N_\text{junc} = L/(L_\text{cnt}/\sqrt{2})$ assuming a saw-type 1D chain of length $L$ made of CNT rodes with crossing angle of $45^{\circ}$. The number of parallel percolation paths over width $W$ can be estimated as $N_\text{path} = W/(L_\text{cnt}/\sqrt{2})$ and the number of 2D layers of CNT monolayers in a film of thickness $t$ can be estimated as $N_\text{layer} = t/(d+h)$. If each junction contributes a conductance of $G_\text{junc}$, then the total conductance due to the junctions in the film should be:
\begin{equation}
    G_\text{film, junc} =N_\text{layer}\frac{N_\text{path}}{N_\text{junc}}  G_\text{inter}=\frac{tW}{L}\frac{1}{(d+h)}G_\text{inter}
\end{equation}
If we consider the conductance contribution from the nanotubes themselves (i.e., the intratube conductance), we can estimate it as follows:
\begin{equation}
    G_\text{film, intra} = N_\text{layer}N_\text{path}\frac{\sigma_\text{intra}}{\sqrt{2}L} =\frac{tW}{L}\frac{1}{(d+h)L_\text{cnt}}\sigma_\text{intra}
\end{equation}
where $\sqrt{2}L$ is the average distance that electron travels in the random film to reach from one electrode to the other and $\sigma_\text{intra}$ is the intratube conductivity.
The net resistance of the 2D film can be estimated as:\cite{shim2015optimally}
\begin{eqnarray}
    G_\text{film}^{-1} &=& G_\text{film, intra}^{-1}+ G_\text{film, inter}^{-1}
    \nonumber \\
    &=&\frac{L}{tW}\,(d+h)\left (\frac{1}{G_\text{inter}}+\frac{L_\text{cnt}}{\sigma_\text{intra}} \right )
\end{eqnarray}
which leads to an estimate of the $C=L(d+h)/(tW)$ value in Eq.~(\ref{eq:Gfilm}). Note that the $C$ value in this simple model does not depend on the CNT volume density as opposed to more sophisticated models~\cite{Komori1977,Shim2015,Tripthy2000}. In our model, CNT volume density is given by $n_{\rm CNT}=N_\text{layer}N_\text{path}N_\text{junc}/(tWL)=2/(L^2_{\rm cnt}(d+h))$ consistent with an estimate in Ref.~\cite{GaoetAl21Carbon}.

\paragraph{\label{Theory:Gel}Electronic Intertube Junction Conductance:} The electronic conductance can be evaluated by~\cite{Dattabook}
\begin{equation}\label{G2}
G_\text{inter, el}=\frac{4\pi e}{\hbar V}\sum_{\substack{\mathbf{k}ss'}}M_{\mathbf{k}s}^{s'}(f(E_{\mathbf{k}s})-f(E_{\mathbf{k}s'}+eV)),
\end{equation}
where the spin degeneracy is included. Here, $\mathbf{k}$ is the 2D electron wave vector in 2D Brillouin zone corresponding to periodically repeated supercell containing a single junction.  $s(s')$ labels the band index, $e$, $\hbar$, $V$, and $f$ are the elementary electronic charge, the reduced Plank constant, the applied voltage bias between layers, and the Fermi-Dirac distribution function, respectively. The intertube coupling $M_{\mathbf{k}s}^{s'}$ is given by
\begin{equation}
M_{\mathbf{k}s}^{s'}= |\mel{\psi_{\mathbf{k}s}}{H_\text{el}}{\psi_{\mathbf{k}s'}}|^2\delta(E_{\mathbf{k}s'}-E_{\mathbf{k}s}),
\end{equation}
where $\mel{\psi_{\mathbf{k}s}}{H_{el}}{\psi_{\mathbf{k}s'}}$ is the tunneling matrix element for carrier scattering from state $\psi_{\mathbf{k}s}$ to state $\psi_{\mathbf{k}s'}$ on different CNTs. $E_{\mathbf{k}s}$ is the carrier energy, calculated by the tight-binding method with the first nearest-neighbor hoping $t=3.1$\,eV.
$H_\text{el}$ is the electron intertube interaction Hamiltonian due to the hopping between atoms on different CNTs. The hopping parameter is defined as \citep{perebeinos2012phonon}
\begin{equation}\label{eq:t_ij}
    t_{ij}=t_{\perp}\exp(-\frac{r_{ij}-h_0}{\lambda_z})\exp[-\left (\dfrac{\xi_{ij}}{\lambda_{xy}} \right )^{\alpha}],
\end{equation}
where $h_0=3.35$\,\AA \ is the equilibrium interlayer distance, $r_{ij}$ is the distance between atoms $i$ and $j$, $\lambda_z=0.6$\,\AA, $\lambda_{xy}=1.7$\,\AA, $\alpha=2.0$, and $t_{\perp}=0.4$\,eV.
$\boldsymbol{\xi}_{ij}$ is the projection of $\mathbf{r}_{ij}$ onto the layers' bisector, and its absolute value is given by
\begin{equation}\label{abs_xi}
   \xi_{ij}=r_{ij}\sin\phi=r_{ij}\sqrt{1- \left (\dfrac{\boldsymbol{\pi}_{ij}\cdot \mathbf{r}_{ij}}{\pi_{ij}r_{ij}} \right )^2} \,.
\end{equation}
$\phi$ is the angle between $\mathbf{r}_{ij}$ and $\boldsymbol{\pi}_{ij}$, $\boldsymbol{\pi}_{ij}=\boldsymbol{\pi}_{i}-\boldsymbol{\pi}_{j}$, where  $\boldsymbol{\pi}_{i}$ is determined by \citep{perebeinos2009valence}
\begin{equation}\label{pi}
   \boldsymbol{\pi}=3\dfrac{\mathbf{r}_{ij}\times\mathbf{r}_{ik}+\mathbf{r}_{ik}\times\mathbf{r}_{il}+\mathbf{r}_{il}\times\mathbf{r}_{ij}}{r_{ij}r_{ik}+r_{ik}r_{il}+r_{il}r_{ij}},
\end{equation}
where $j$, $k$, and $l$ are the three nearest neighbors of $i$. $\boldsymbol{\pi}_i$ is perpendicular to the surface of the CNT, and its direction points toward the outside of CNT, and $\boldsymbol{\xi}_{ij}$ is perpendicular to $\boldsymbol{\pi}_{ij}$.

\paragraph{\label{Theory:Gel}Phonon-Assisted Intertube Junction Conductance:}The phonon-assisted conductance is computed by \citep{perebeinos2012phonon}
\begin{equation}
G_\text{inter, ph}=\frac{4\pi e}{\hbar V}\sum_{\substack{\mathbf{k} ,\mathbf{k'},s,s',\mu}}M_{\mathbf{k'}s'\mu}^{\mathbf{k}s}[f(E_{\mathbf{k}s})-f(E_{\mathbf{k'}s'}+eV)].
\label{eq:Iph}
\end{equation}
Here, $M_{\mathbf{k'}s'\mu}^{\mathbf{k}s}$ is given by
\begin{equation}
\begin{aligned}
M_{\mathbf{k'}s'\mu}^{\mathbf{k}s}={}& |\mel{\psi_{\mathbf{k}s}}{H_\text{e-ph}^{\mu}}{\psi_{\mathbf{k'}s'}}|^2[n_{\mathbf{q}\mu}\delta(E_{\mathbf{k'}s'}-E_{\mathbf{k}s}+\hbar\omega_{\mathbf{q}\mu}) \\
& (1+n_{-\mathbf{q}\mu})\delta(E_{\mathbf{k'}s'}-E_{\mathbf{k}s}-\hbar\omega_{-\mathbf{q}\mu})],
\end{aligned}
\end{equation}
where $n$ is the Bose-Einstein function, $\mel{\psi_{\mathbf{k}s}}{H_\text{e-ph}^{\mu}}{\psi_{\mathbf{k'}s'}}$ is the electron-phonon matrix element, and $\omega_{\mathbf{q}\mu}$ is the phonon frequency with wave vector $\mathbf{q}=\mathbf{k}-\mathbf{k'}$ and branch number $\mu$.
To obtain the phonon frequency, the structure is relaxed using the BFGS algorithm of the quasi-Newton method.\cite{avriel2003nonlinear}
The electron-phonon Hamiltonian are extracted by expanding Equation~\eqref{eq:t_ij} in atomic displacements corresponding to the phonon normal modes.

\begin{acknowledgements}

We gratefully acknowledge funding from the Vice President for Research and Economic Development (VPRED), SUNY Research Seed Grant Program, and computational facilities at the Center for Computational Research at the University at Buffalo (\url{http://hdl.handle.net/10477/79221}).
A.M.\ and J.K.\ acknowledge support by the Basic Energy Science (BES) program of the U.S.\ Department of Energy through Grant No.\ DE-FG02-06ER46308 (for preparation of carbon nanotube films) and the Robert A.\ Welch Foundation through Grant No.\ C-1509 (for structural and electrical characterization measurements).

\end{acknowledgements}

%\bibliographystyle{achemso}
\bibliography{References}

\end{document}